\documentclass[journal]{IEEEtran}   
\usepackage{multirow}
\usepackage{diagbox}
\usepackage{amssymb}
\usepackage{amsmath}
\usepackage{graphicx}
\usepackage{cite}
\usepackage{citesort}
\usepackage{subfigure}
\usepackage{graphicx,epstopdf}
\usepackage{epsfig}	
\usepackage{cite,graphicx,amsmath,amssymb}
\usepackage{comment}
\makeatletter
\renewcommand{\maketag@@@}[1]{\hbox{\m@th\normalsize\normalfont#1}}%
\makeatother
\usepackage{caption}
\usepackage{algorithm}
\usepackage{algorithmic}
\usepackage{amsmath} 
\usepackage{amssymb}
\usepackage{amsmath}
\usepackage{cite}
\usepackage{url}
\usepackage{amsmath} 
\usepackage{bm}      
\usepackage{xcolor}
\usepackage{cite,graphicx,amsmath,amssymb}
\usepackage{subfigure}
\usepackage{citesort}
\usepackage{fancyhdr}
\usepackage{mdwmath}
\usepackage{mdwtab}
\usepackage{caption}
\usepackage{amsthm}
\usepackage{lipsum}
\usepackage{subcaption}

\newtheorem{theorem}{Theorem}

\newtheorem{lemma}{Lemma}

\newtheorem{corollary}{Corollary}

\makeatletter
\def\ScaleIfNeeded{%
\ifdim\Gin@nat@width>\linewidth \linewidth \else \Gin@nat@width
\fi } \makeatother

\begin{document}

\title{\Huge{Fluid Antennas Assisted RIS-NOMA Communication Networks}}

\author{ 
 Xinwei Yue,~\IEEEmembership{Senior Member,~IEEE}, He Geng, Jingjing Zhao,~\IEEEmembership{Member,~IEEE},\\
 Xianli Gong,~\IEEEmembership{Member,~IEEE}, Aryan Kaushik,~\IEEEmembership{Member,~IEEE}, Arumugam Nallanathan,~\IEEEmembership{Fellow,~IEEE}

\thanks{X. Yue and H. Geng are with the College of Information and Communication Engineering, and with the Center for Target Cognition Information Processing Science and Technology, Beijing Information Science $\&$ Technology University, and also with the Key Laboratory of Modern Measurement $\&$ Control Technology, Ministry of Education, Beijing Information Science $\&$ Technology University, Beijing 102206, China (email: \{xinwei.yue and he.geng\}@bistu.edu.cn).}
\thanks{J. Zhao is with the School of Electronic and Information Engineering, Beihang University, Beijing 100191, China (email: jingjingzhao@buaa.edu.cn).}
\thanks{X. Gong is with the School of Information Engineering, North China University of Water Resources and Electric Power, Zhengzhou 450046, China (e-mail: gongxianli@ncwu.edu.cn).}
\thanks{A. Kaushik is with the Department of Computing and Mathematics, Manchester Metropolitan University, UK (e-mail: a.kaushik@ieee.org).}
\thanks{A. Nallanathan is with the School of Electronic Engineering and Computer Science, Queen Mary University of London, London E1 4NS, U.K. (email: a.nallanathan@qmul.ac.uk).}
 }

\maketitle

\begin{abstract}
This paper introduces a fluid antenna system (FAS) into reconfigurable intelligent surface (RIS) assisted non-orthogonal multiple access (NOMA) communication networks, where the non-orthogonal users are equipped with planar fluid antennas. Specifically, we formulate a sum rate maximization problem for FAS-RIS-NOMA networks, which jointly optimizes the fluid ports, the RIS deployment, and the phase shift matrix. To solve the resulting non-convex optimization problem involving highly-coupled variables, an iterative algorithm based on alternating optimization is employed to decompose the original problem into three subproblems. Exhaustive search is employed for the optimal fluid ports, particle swarm optimization for the RIS deployment, and semidefinite relaxation with successive convex approximation for the optimal phase shift matrix, respectively. Finally, the presented simulation results show that: 1) Compared to traditional antenna systems and orthogonal multiple access, the FAS-RIS-NOMA networks achieves the larger system throughput under the high signal-to-noise ratio conditions; 2) By increasing the number of RIS elements and enlarging the FAS size, the sum rate of FAS-RIS-NOMA networks is enhanced significantly.

\end{abstract}
\section{Introduction}

With the number of device in wireless networks explosively grows, the sixth generation of wireless networks will face a thousand times the speed requirements and massive connection demands of the fifth generation\cite{To6G1,To6G2}. Improving link quality within limited physical space has become a key challenge. Fluid antenna system (FAS) has ability to solve this problem by using movable metal or liquid conductive media inside flexible tubes to dynamically switch and optimize antenna ports\cite{2021FAS,2024tutorialfluidantenna6g}. Furthermore, the Gaussian copula approach of FAS demonstrated that increasing the fluid channel size and the number of ports is capable of reducing the system outage probability\cite{GaussianCopula}. 

Reconfigurable intelligent surfaces (RIS) garnered significant attention for their dynamic environmental manipulation \cite{2021liuRIS,xiejin}. Building on this, combining RIS with FAS leveraged RIS phase shifts to proactively optimize channels, enhancing data rates and outage probabilities \cite{2024RISFAS}. Extending to multi-user scenarios, FAS-RIS systems analyzed sum rate and secrecy outage, validating the advantages of RIS in enhancing security performance \cite{2025FASRIS}.
To tackle complexity arising from such deployments, the authors of \cite{zhang2025fluidantennameetsris} proposed a dual-timescale design that reduced real-time overhead in FAS-RIS systems. Furthermore, FAS-RIS exploited the dynamic adjustment capability of movable elements to elevate weighted sum rate  \cite{zhao2024movables}. Until now, FAS has been introduced into NOMA networks by overcoming multipath propagation. The authors of \cite{202501NOMAFAS} surveyed the superiority of FAS-NOMA networks in terms of outage behaviors. 
By taking into account imperfect channel state information (ipCSI), the authors in \cite{2024FASNOMAOMA} confirmed that the enhanced gains of FAS-NOMA is achieved  via optimal port selection. Considering the aforementioned advantages, the authors of \cite{2025NOMARISFAS} analyzed the outage probability of  FAS-RIS enabled NOMA non-terrestrial networks. 

In scenarios with blocked direct links and multiple coexisting users, a single technique is generally insufficient to simultaneously guarantee reliable link enhancement and high spectral efficiency. Although RIS can reconfigure the propagation environment through programmable reflections, it lacks adaptive spatial sampling capability at the receiver side. The ability of FAS is to exploit port selection to harvest additional spatial degrees of freedom and diversity gains, yet it alone cannot effectively compensate for the severe cascaded-channel attenuation caused by blockage. In addition, NOMA is capable of improving spectrum utilization, but its performance strongly depends on channel disparity among users and the overall link quality. To address the above challenges in a unified manner, this paper integrates FAS technique into RIS-NOMA networks for more efficient multi-user transmission in blockage-prone environments.
Specifically, we formulate a non-convex optimization problem to maximize the sum rate of FAS-RIS-NOMA networks. To solve this, we propose a unified framework that jointly optimizes the fluid antenna port selection, the 3D RIS deployment, and the RIS phase shifts, dividing the main problem of maximizing the sum rate of the FAS-RIS-NOMA networks into three subproblems. An exhaustive search is used to determine the optimal port, particle swarm optimization (PSO) is applied for RIS location optimization and semidefinite relaxation (SDR) along with successive convex approximation (SCA) methods are used for optimizing the RIS phase shift matrix. The process is iterated until convergence to efficiently obtain the global optimal solution. Finally, the experimental results show that the FAS-RIS-NOMA networks achieve the larger system throughput compared to traditional antenna systems (TAS) and orthogonal multiple access (OMA) in the high signal-to-noise (SNR) region. Increasing the number of RIS elements and enlarging the FAS size, the sum rate of FAS-RIS-NOMA networks is enhanced significantly.    
\section{System Model}\label{SYSTEM AND CHANNEL MODELS}
\subsection{Network Model}
We consider the FAS-RIS-NOMA communication scenarios, where a base station (BS) equipped with $L$ antennas transmits superimposed signals to a nearby user $\mathrm{U}_n$ and distant user $\mathrm{U}_m$. Users employ two dimensional (2D) fluid antennas. Each antenna ports are distributed in a grid configuration across a planar region with a total of $K$ ports. The arrangement consists of $K_1$ elements along the length and $K_2$ along the width, i.e., $K = K_1 \times K_2$. The array covers a normalized area of dimensions $W = W_1\lambda \times W_2\lambda$, where $W_1$ and $W_2$ denote normalized sizes and $\lambda$ represents the carrier wavelength\cite{2023FASIII}. In addition, we provide a mapping function $f(k)=(k_{1},k_{2})$ and $f^{-1}(k_{1},k_{2})=k$ in order to transform the 2D into a one- dimensional index, where $k\in\{1,\cdots,K\}$, $k_{1}\in\{1,\cdots,K_{1}\}$ and $k_{2}\in\{1,\cdots,K_{2}\}$. For clarity of exposition,  we assume that the direct links are obstructed by obstacles and a UAV-mounted RIS is considered as a relay node to establish additional communication links. Three dimensional (3D) Cartesian coordinate system is employed to describe the locations of communication nodes. The coordinates of the BS, $\mathrm{U}_n$ and $\mathrm{U}_m$ are respectively expressed as \(\mathbf{q}_b = [x_{b},y_{b},z_{b}]^{T}\), \(\mathbf{q}_n = [x_{n},y_{n},z_{n}]^{T}\) and \(\mathbf{q}_m=[x_{m},y_{m},z_{m}]^{T}\). RIS is positioned at \(\mathbf{q}_r = [x_r, y_r, z_r]^T\) and comprises \(M = M_1 \times M_2\) passive reflecting elements, where \(M_1\) and \(M_2\) denote the number of elements along the \(x\)-axis and \(y\)-axis, respectively.

\subsection{Signal Model}
The signal received by $\mathrm{U}_i$, $\forall i \in \{n, m\}$, can be expressed as
\begin{equation}
\boldsymbol{y_i} = \mathbf{H}_{br}\boldsymbol{\Theta}\mathbf{H}_{ri} \left( \sqrt{a_n P_s x_n} + \sqrt{a_m P_s x_m} \right) +\boldsymbol{{z}_{i}},
\end{equation}
where $ x_n $ and $ x_m $ are the signals of $\mathrm{U}_n$ and $\mathrm{U}_m$, respectively. $ P_s $ is the normalized transmission power of the BS. $ a_n $ and $ a_m $ are the corresponding power allocation coefficients which satisfies the condition $ a_m > a_n $ and $ a_n + a_m = 1 $ to ensure the better fairness. The reflection coefficient matrix of RIS is defined as \(\boldsymbol{\Theta}=\text{diag}(e^{j\theta_{1}},e^{j\theta_{2}},\cdots,e^{j\theta_{M}})\in\mathbb{C}^{M\times M}\), $\theta_{m}\in[0, 2\pi)$ represents the reflection phase shift of the mth RIS. $\boldsymbol{{z}_{i}} \sim{\cal C}{\cal N} \left( {0,\sigma _0^2\boldsymbol{I}} \right)$ stands for white Gaussian noise at position $\mathrm{U}_i$ with average power ${\sigma _0^2\boldsymbol{I}}$, where $\boldsymbol{I}$ is the identity matrix. We represent the channel coefficient from the BS to the RIS as \( \mathbf{H}_{br} \in \mathbb{C}^{L \times M} \). The corresponding expression for $\mathbf{H}_{br}$ can be given by
 $\mathbf{H}_{br} = \sqrt{\rho_0 d_{br}^{-\alpha_0}} \left( \sqrt{\frac{\kappa_{br}}{\kappa_{br}+1}} \bm{\bar{\mathbf{H}}}_{br} + \sqrt{\frac{1}{\kappa_{br}+1}} \bm{\tilde{\mathbf{H}}}_{br} \right),$
where $\rho_0$ indicates path loss per unit distance, $\alpha_0$ is the path loss exponent, and $d_{br} = \| \bm{q}_b - \bm{q}_r \|$ represents the Euclidean distance between the BS and RIS. $\kappa$ denotes the Rice factor. The line of sight (LoS) component is expressed as $\bm{\bar{\mathbf{H}}}_{br} = \bm{\alpha}_N \bm{\alpha}_R^H$, where $\bm{\alpha}_N$ and $\bm{\alpha}_R$ are the steering vectors for the ULA at the BS and the UPA at the RIS, respectively. These vectors are defined as 
$\bm{\alpha}_N = \left[ 1, \cdots, e^{-j \frac{2\pi}{\lambda} d(L-1) \cos \varphi} \right]^T$ and 
$\bm{\alpha}_R = \left[ 1, \cdots, e^{-j \frac{2\pi}{\lambda} d_1 (M_1-1) \sin \varphi_{br} \cos \phi_{br}} \right]^T 
\otimes \left[ 1, \cdots, e^{-j \frac{2\pi}{\lambda} d_2 (M_2-1) \cos \varphi_{br}} \right]^T$, where $d$ is the antenna spacing at the BS, and \(d_1\), \(d_2\) denote the element spacing along the \(x\)-axis and \(y\)-axis of the RIS, respectively. We separately define $\phi_{br}$ and $\varphi_{br}$ as the azimuth angle of arrival (AOA) and the elevation AOA from BS to RIS. $\varphi$ is defined as the azimuth angle-of-departure (AOD) from the BS. At this point, the relative positions of each node can be defined as $\sin \varphi_{br} \cos \phi_{br} = \frac{|x_b - x_r|}{d_{br}}, \cos \varphi_{br} = \frac{|z_b - z_r|}{d_{br}}, \cos \varphi = \frac{|y_b - y_r|}{d_{br}}$. The no line of sight (NLOS) component follows a complex Gaussian distribution $\mathbf{\tilde{H}}_{br} \sim \mathcal{CN}(0, 1)$.

The channel from the RIS to $\mathrm{U}_i$ is represented as \( \mathbf{H}_{ri} \in \mathbb{C}^{M \times K} \), which we model as
 $\mathbf{H}_{ri} = \sqrt{\rho_0 d_{ri}^{-\alpha_0}} \left(\sqrt{\frac{\kappa_i}{\kappa_i + 1}} \mathbf{\overline{H}} _{ri} + \sqrt{\frac{1}{\kappa_i + 1}} \mathbf{\tilde{H}}_{ri}\mathbf{R_i}^{\frac{1}{2}}\right)$, where the Euclidean distance between the RIS and $\mathrm{U}_i$ is represented by $d_{ri} = \left\| \mathbf{q}_r - \mathbf{q}_i \right\|$. The LoS component is expressed as $\overline{\mathbf{H}}_{ri} = \mathbf{a}_{\mathrm{T},i} \mathbf{a}_{\mathrm{U},i}^\mathrm{H} $, where the RIS steering vector $\mathbf{a}_\mathrm{U}$ and $\mathbf{a}_\mathrm{T}$ are defined as the Kronecker product of the horizontal and vertical steering vectors i.e., 
$\mathbf{a}_{\mathrm{U,i}} = 
\begin{bmatrix}
1,\ e^{j 2\pi W_1 \sin\varphi_{ri} \cos\phi_{ri} / (K_1 - 1)},\ \dots,\ e^{j 2\pi W_1 \sin\varphi_{ri} \cos\phi_{ri}}
\end{bmatrix}^T 
\otimes 
\begin{bmatrix}
1,\ e^{j 2\pi W_2 \sin\varphi_{ri} \sin\phi_{ri} / (K_2 - 1)},\ \dots,\ e^{j 2\pi W_2 \sin\varphi_{ri} \sin\phi_{ri}}
\end{bmatrix}^T$ 
and 
$\mathbf{a}_\mathrm{T,i} = \begin{bmatrix}
1, \dots, e^{-j \frac{2 \pi}{\lambda} d_1 (M_1 - 1) \sin\varphi_{ri} \cos\phi_{ri}}
\end{bmatrix} \otimes \begin{bmatrix}
1, \dots, e^{-j \frac{2 \pi}{\lambda} d_2 (M_2 - 1) \cos\varphi_{ri}}
\end{bmatrix}$. 
We define $\sin \varphi_{ri} \cos \phi_{ri} = \frac{|x_{i} - x_r|}{d_{ri}}$, $\sin \varphi_{ri} \sin \phi_{ri} = \frac{|y_{i} - y_r|}{d_{ri}}$, and $\cos \varphi_{ri} = \frac{|z_{i} - z_r|}{d_{ri}}$, where $\phi_{ri}$ and $\varphi_{ri}$ are the azimuth and elevation AoA from the RIS to user $\mathrm{U}_i$.
The NLoS component follows a complex Gaussian distribution $\mathbf{\tilde{H}}_{ri} \sim \mathcal{CN}(0, 1)$. 
Note that due to the close proximity of the ports in the FAS, the corresponding channels are spatially correlated. We adopt the Jakes' model, where the correlation coefficient between the $k$-th port and the $\tilde{k}$-th port is modeled by\cite{MIMO_FAS}
\begin{equation}
\rho_{k, \tilde{k}} =J_0 \left( 2\pi \sqrt{\left(\frac{k_1 - \tilde{k}_1}{K_1 - 1} W_1\right)^2 + \left(\frac{k_2 - \tilde{k}_2}{K_2 - 1} W_2\right)^2} \right),
\end{equation}
where $J_0(\cdot)$ is the zero-order Bessel function of the first kind. Therefore, the spatial correlation matrix $\mathbf{R_i}$ at $\mathrm{U}_i$ is defined as\cite{wu2026tamingbessellandscapejoint}
\begin{equation}
\mathbf{R_i} = 
\begin{bmatrix}
\rho_{1,1} & \rho_{1,2} & \cdots & \rho_{1,K} \\
\rho_{2,1} & \rho_{2,2} & \cdots & \rho_{2,K} \\
\vdots & \vdots & \ddots & \vdots \\
\rho_{K,1} & \rho_{K,2} & \cdots & \rho_{K,K}
\end{bmatrix}.
\end{equation}
At the receiver $\mathrm{U}_n$, it first decodes $\mathrm{U}_m$ and employs SIC to eliminate the interference before decoding its own signal. During this process, the SNR of detecting $\mathrm{U}_m$ at $\mathrm{U}_n$ is expressed as 
\begin{equation} \setlength\belowdisplayskip{3pt}
\gamma_{\mathrm{U}_m \to \mathrm{U}_n} = \frac{\| \mathbf{H}_{br} \Theta \mathbf{H}_{rn} \|^2 a_m \rho}{\| \mathbf{H}_{br} \Theta \mathbf{H}_{rn} \|^2 a_n \rho + 1},
\end{equation}
where the parameter $\rho$ is the transmit SNR, defined as $\rho = \frac{P_s}{\sigma _0^2}$.
After completing the SIC for $\mathrm{U}_m$, $\mathrm{U}_n$ decodes its own signal. Since the interference from $\mathrm{U}_m$ has been removed, the SNR for $\mathrm{U}_n$ to detect its own information can be expressed as\cite{songxinlong} 
\begin{equation}\setlength\belowdisplayskip{3pt}
\gamma_{\mathrm{U}_n} = \| \mathbf{H}_{br} \boldsymbol{\Theta} \mathbf{H}_{rn} \|^2 a_n \rho.
\end{equation}
$\mathrm{U}_m$ receives the signal while being interfered with by the signal from $\mathrm{U}_n$. The SNR for $\mathrm{U}_m$ to detect its own signal can be expressed as 
\begin{equation}\setlength\belowdisplayskip{3pt}
\gamma_{\mathrm{U}_m} = \frac{\| \mathbf{H}_{br} \boldsymbol{\Theta} \mathbf{H}_{rm} \|^2 a_m \rho}{\| \mathbf{H}_{br} \boldsymbol{\Theta} \mathbf{H}_{rm} \|^2 a_n \rho + 1}.
\end{equation}
\section{Maximize Sum Rate Strategy}\label{MAXIMIZE TOTAL RATE STRATEGY}
This section aims to maximize the sum rate of FAS-RIS-NOMA by jointly optimizing the port selection of FAS and the deployment locations and phase shift matrices of the RIS. We define the main problem \textbf{P1} as follows 
\begin{eqnarray}
\begin{split}
\textbf{P1:} \max\limits_{\mathbf{k_i}, \mathbf{q}_r, \boldsymbol{\Theta}}  \quad R &= \log \left( 1 + \left\| \mathbf{H}_{br} \boldsymbol{\Theta}  \mathbf{H}_{rn} \right\|^2 a_n \rho \right) \\
  &\quad + \log \left( 1 + \frac{\left\| \mathbf{H}_{br} \boldsymbol{\Theta} \mathbf{H}_{rm} \right\|^2 a_m \rho}{\left\| \mathbf{H}_{br} \boldsymbol{\Theta} \mathbf{H}_{rm} \right\|^2 a_n \rho + 1} \right)
\end{split}
\end{eqnarray}
\begin{equation*}
\begin{array}{ll}
\hspace{-1.5cm}
\quad\quad\text{s.t.} 
 & C1:a_n + a_m = 1, \\
 & C2:R \geq R_{\min}, \\
 & C3:0 \leq \theta_m \leq 2\pi, \\
 & C4: \ \|f(k)-f(\tilde{k})\| \geq d_0,\quad k \neq \tilde{k},\\
 & C5:\mathbf{q}_r \in \mathrm{U},\\
\end{array}
\end{equation*}
where the constraint (C1) ensures that the power allocation coefficients for the two users sum to unity, maintaining total power normalization. The constraint (C2) guarantees that the system sum rate meets or exceeds the minimum required rate $R_{\text{min}}$. The constraint (C3) restricts the phase shifts of the RIS elements to valid angular ranges for proper signal alignment. The constraint (C4) specifies that the distance between any two different candidate fluid antenna ports should not be smaller than \(d_0\). Here, \(k\) and \(\tilde{k}\) are port indices, while \(f(k)\) and \(f(\tilde{k})\) represent their corresponding 2D port positions. Finally, the constraint (C5) confines the spatial coordinates of the RIS to predefined ranges, ensuring feasible deployment within the system physical boundaries.

Due to variable coupling, the main problem is hard to solve. To address this challenge, we will decompose the overall problem into three subproblems using the AO algorithm, i.e., optimizing the FAS port selection, the deployment locations and phase-shift matrices of the RIS.
\subsection{Fluid Antenna Port Optimization SubProblem}
In this subsection, the subproblem optimizes the fluid antenna port selection of two users for a fixed RIS deployment location $\mathbf{q}_r$ and phase shift matrix $\mathbf{\Theta}$ thereby maximizing the sum rate of the FAS-RIS-NOMA networks. The subproblem is thus formulated as
\begin{equation} \label{eq:P1.1}
\begin{aligned}
\textbf{P1.1: } \max\limits_{\mathbf{k_i^*}} \quad R &= \log \left( 1 + \left\| \mathbf{H}_{br} \boldsymbol{\Theta}  \mathbf{H}_{rn} \right\|^2 a_n \rho \right) \\
  &\quad + \log \left( 1 + \frac{\left\| \mathbf{h}_{br} \boldsymbol{\Theta} \mathbf{H}_{rm} \right\|^2 a_m \rho}{\left\| \mathbf{H}_{br} \boldsymbol{\Theta} \mathbf{H}_{rm} \right\|^2 a_n \rho + 1} \right) \\
  \text{s.t.} \quad &C4: \ \|f(k)-f(\tilde{k})\| \geq d_0,\quad k \neq \tilde{k}.\\
\end{aligned}
\end{equation}
Given the limited number of ports in the FAS, this subproblem is efficiently solved using exhaustive search. Specifically, for each $\mathrm{U}_i$, we compute the equivalent channel gain for all $K$ ports and select the port index that yields the maximum gain through one dimensional search, the port of maximum gain at this point is denoted as
\begin{equation}
\mathbf{k_i^*} = \underset{k \in \{1, 2, \ldots, K\}}{\arg\max} \left\| \mathbf{h}_{br}\mathbf{\Theta}\mathbf{h}_{ri, k} \right\|^2.
\end{equation}
\subsection{RIS Deployment Location Optimization SubProblem}
The core of this subproblem lies in determining the optimal 3D coordinates $\mathbf{q}_r$ for the RIS. The RIS deployment location directly influences key parameters such as the channel gain between users and BS, signal transmission path loss, and signal arrival angles, thereby significantly affecting the sum rate of the FAS-RIS-NOMA networks. Due to the complex nonlinear relationship between the RIS deployment location and the sum rate, as well as the dynamic nature of the channel environment, this is a typical non-convex optimization problem, which traditional convex optimization methods struggle to solve effectively.
We define the RIS deployment location optimization subproblem as follows:
\begin{equation} \label{eq:P1.2}
\begin{aligned}
\textbf{P1.2: } \max\limits_{\mathbf{q}_r^*} \quad R &= \log \left( 1 + \left\| \mathbf{H}_{br} \boldsymbol{\Theta}  \mathbf{H}_{rn} \right\|^2 a_n \rho \right) \\
  &\quad + \log \left( 1 + \frac{\left\| \mathbf{H}_{br} \boldsymbol{\Theta} \mathbf{H}_{rm} \right\|^2 a_m \rho}{\left\| \mathbf{H}_{br} \boldsymbol{\Theta} \mathbf{H}_{rm} \right\|^2 a_n \rho + 1} \right) \\
\quad \text{s.t.} \quad &  \mathbf{q}_r \in \mathrm{U}.
\end{aligned}
\end{equation}

To address this challenge, the PSO algorithm is employed to optimize the deployment locations of RIS. During the algorithm's iterative process, each particle represents a potential deployment location solution. The particles update their velocity and position based on their personal best position $p_{\text{best}}$ and the global best position $g_{\text{best}}$ found by the swarm.

Specifically, the velocity $v_\gamma(t)$ and position $q_\gamma(t)$ update equations for particle $\gamma$ at iteration $t$ are given by
\begin{equation} \label{eq:pso_update_combined}
\begin{aligned}
v_\gamma(t+1) &= \omega v_\gamma(t) + c_1 r_1 (p_{best}(t) - q_\gamma(t)) \\
&\quad + c_2 r_2 (g_{best}(t) - q_\gamma(t)), \\
q_\gamma(t+1) &= q_\gamma(t) + v_\gamma(t+1),
\end{aligned}
\end{equation}
where $\omega$ is the inertia weight, which balances the global and local search capabilities. $c_1$ and $c_2$ are the learning factors, controlling the particle's tendency to move towards its personal best and the global best, respectively. $r_1$ and $r_2$ are random numbers uniformly distributed within the range $[0, 1]$. The iterative nature of the particle swarm optimization algorithm enables it to efficiently locate global optima within multidimensional non-convex search spaces, thereby avoiding getting stuck in local optima. The algorithm will continue to iterate until the maximum number of iterations is reached or when the change in the global optimal solution is less than a threshold, and finally outputs the optimal RIS deployment location $\mathbf{q}_r^*$.

\subsection{Reflecting Phase Shift Optimization Subproblem}
The original optimization problem exhibits inherent non-convexity due to constant modulus phase constraints and multi-user interference components. To address this challenge, a joint optimization framework based on SDR and SCA is proposed in this study.
To address the non-convexity arising from the unit modulus constraints, the RIS phase vector $\mathbf{v}=[e^{j\theta_1}, \dots, e^{j\theta_M}]^T$ is lifted to the semidefinite matrix space. By constructing $\mathbf{V} = \mathbf{v}\mathbf{v}^H$, the unit modulus constraint is equivalently transformed into \(\mathbf{V}_{mm} = 1\). However, this transformation introduces a non-convex rank-one constraint, i.e., \(\operatorname{rank}(\mathbf{V}) = 1\). To render the problem tractable, this rank constraint is relaxed, thereby converting the original non-convex optimization problem into a semi-definite program. At this point, subproblem is expressed as
\begin{equation} \label{eq:P1.3}
\begin{aligned}
\textbf{P1.3: } \max_{\mathbf{V}} \; & \quad R=\log \left(1 + a_n \rho \operatorname{Tr}(\hat{\mathbf{H}}_{br} \mathbf{V} \hat{\mathbf{H}}_{rn})\right) \\
&\quad + \log \left(1 + \frac{a_m \rho \operatorname{Tr}(\hat{\mathbf{H}}_{br} \mathbf{V} \hat{\mathbf{H}}_{rm})}{a_n \rho \operatorname{Tr}(\hat{\mathbf{H}}_{br} \mathbf{V} \hat{\mathbf{H}}_{rm}) + 1}\right) \\
\quad \text{s.t.} \quad & \mathbf{V}_{mm} = 1, \quad \forall m \in \{1, \dots, M\},
\end{aligned}
\end{equation}
where \(\hat{\mathbf{H}}_{br} = \mathbf{H}_{br} \mathbf{H}_{br}^H\) represents the channel covariance matrix from the BS to the RIS, while \(\hat{\mathbf{H}}_{ri} = \mathbf{H}_{ri} \mathbf{H}_{ri}^H\) denotes the channel covariance matrix from the RIS to $\mathrm{U}_i$. Through this SDP transformation, the composite channel gain \(\| \mathbf{H}_{br} \mathbf{\Theta} \mathbf{H}_{ri} \|^2\) is precisely expressed as the matrix trace operation \(\operatorname{Tr}(\hat{\mathbf{H}}_{br} \mathbf{V} \hat{\mathbf{H}}_{ri})\).

Despite SDR having resolved the unit modulus constraint, the fractional term \(\displaystyle \frac{a_m \rho\,\operatorname{Tr}(\hat{\mathbf{H}}_{br}\mathbf{V}\hat{\mathbf{H}}_{rm})}{a_n \rho\,\operatorname{Tr}(\hat{\mathbf{H}}_{br}\mathbf{V}\hat{\mathbf{H}}_{rm}) + 1}\) 
in the objective still introduces non-convexity. To address this, we introduce the auxiliary variable \(\mu = \operatorname{Tr}(\hat{\mathbf{H}}_{br}\mathbf{V}\hat{\mathbf{H}}_{rm})\) and exploit the global concavity of \(\displaystyle f(\mu) = \frac{a_m \rho\,\mu}{a_n \rho\,\mu + 1}\). In each iteration, we perform a first-order Taylor expansion around the current point to obtain a linear approximation of \(f(\mu)\), thereby transforming the original problem into a sequence of standard convex subproblems. Each subproblem can be efficiently solved using CVX, and the process is repeated until convergence to the final solution. However, SDR relaxation typically yields solutions \(\mathbf{V}^*\) that are not rank-one. To address this discrepancy, Gaussian randomization is employed to recover a physically compliant rank-one solution. Specifically, \(S\) random vectors \(\mathbf{u}^{(s)}\) are sampled from the distribution \(\mathcal{CN}(\mathbf{0}, \mathbf{V}^*)\). Subsequently, these vectors are projected onto the unit modulus circle via the phase projection \(\tilde{\mathbf{v}}^{(s)} = e^{j \arg(\mathbf{u}^{(s)})}\), yielding a set of candidate solutions satisfying the physical constraints. The final phase configuration \(\mathbf{v}^*\) is then selected from these candidate solutions based on the maximum sum rate criterion \(\mathbf{v}^* = \arg \max\limits_{\tilde{\mathbf{v}}^{(l)}} R(\tilde{\mathbf{v}}^{(l)})\).
\subsection{Analysis of Approximate Solutions}
Since the original problem is highly coupled and non-convex, and the proposed AO framework involves heuristic search, successive convex approximation, and semidefinite relaxation, it is difficult to establish a strict global optimality guarantee for the overall joint optimization problem. Specifically, Subproblem~1 is solved by exhaustive search under given RIS deployment and phase-shift variables, and therefore its conditionally optimal solution can be obtained. For Subproblem~2, the RIS deployment location is optimized by PSO. This method is well suited for complex non-convex continuous optimization problems and can achieve a high-performance deployment solution with affordable complexity, although its global optimality is generally difficult to establish rigorously~\cite{Zhou2024Heuristic}. Subproblem~3 is handled by combining SDR, SCA, and Gaussian randomization, and its solution is therefore essentially approximate in nature~\cite{Li2021Joint}. To quantify the approximation error in this step, let $R_{\mathrm{app}}$ denote the optimal objective value of the approximated problem after SCA and SDR, and let $R_{\mathrm{feas}}$ denote the objective value of the recovered feasible solution. Then, we have
$
\eta=\frac{R_{\mathrm{app}}-R_{\mathrm{feas}}}{R_{\mathrm{app}}}.
$
Accordingly, the proposed AO algorithm can be regarded as a solution framework consisting of conditional optimality, heuristic search, and approximation-based optimization. Although a strict global optimality conclusion is difficult to establish for the overall problem, the proposed method is still capable of achieving excellent performance with acceptable computational complexity.

\subsection{Algorithmic Complexity}
The overall computational complexity of proposed AO algorithm is $\mathcal{O}\left( I \left( K + S T_{\text{PSO}} LMK + L_{\text{SCA}} M^{4.5} \right) \right)$\cite{guoxinning}, where $I$ denotes the number of outer AO iterations, $S$ and $T_{\text{PSO}}$ represent the population size and maximum number of iterations of the PSO algorithm respectively, and $L_{\text{SCA}}$ indicates the number of SCA iterations. 
\begin{algorithm}[h]	
\small
\caption{Alternating Optimization Algorithm for Deriving $\mathbf{k}^*_i, \mathbf{q}_r^*, \boldsymbol{\Theta}^*$}
\label{alg:AO_algorithm}
\begin{algorithmic}[1]
\REQUIRE System parameters, $I_{\max}$, $\epsilon$
\ENSURE $\mathbf{k}^*_n, \mathbf{k}^*_m, \mathbf{q}_r^*, \boldsymbol{\Theta}^*, R^*$
\STATE Initialize $\mathbf{q}_r^{(0)} \in \mathrm{U}$, $\boldsymbol{\Theta}^{(0)}$, $R^{(0)} = 0$, $I = 0$
\STATE \textbf{while} {$I < I_{\max}$ and $|R^{(I)} - R^{(I-1)}| > \epsilon$}

\STATE For given $\mathbf{q}_r^{(I-1)}$ and $\boldsymbol{\Theta}^{(I-1)}$, solve Problem P1.1 and obtain $\mathbf{{k}^*_i}^{(I)}$;
\STATE For given $\mathbf{{k}^*_i}^{(I)}$ and $\boldsymbol{\Theta}^{(I-1)}$, solve Problem P1.2 and obtain $\mathbf{q}_r^{(I)}$;
\STATE For given $\mathbf{{k}^*_i}^{(I)}$ and $\mathbf{q}_r^{(I)}$, solve Problem P1.3 and obtain $\boldsymbol{\Theta}^{(I+1)}$;
\STATE $I \leftarrow I + 1$
\STATE \textbf{end while}
\RETURN $\mathbf{k}^*_i, \mathbf{q}_r^*, \boldsymbol{\Theta}^*, R^*$
\end{algorithmic}
\end{algorithm}
\section{NUMERICAL RESULTS}\label{NUMERICAL RESULTS} 
In this section, we validate the effectiveness of the proposed AO algorithm through numerical simulations. The scenario is modeled in a 3D Cartesian coordinate system. The BS, RIS, $\mathrm{U}_n$, and $\mathrm{U}_m$ are located at $(0,0,0)\,\mathrm{m}$, $(50,70,30)\,\mathrm{m}$, $(80,50,0)\,\mathrm{m}$, and $(200,20,0)\,\mathrm{m}$, respectively. Unless otherwise stated, we assume that $\kappa_{br}=1$, $\kappa_{ri}=0$, $\alpha_0=2.2$, SNR$=30\,$dB, $M=64$, $W=2\lambda\times1\lambda$ and $K=20\times10$\cite{2024FAS}. Moreover, $\mathrm{U}$ is specified by $x_r \in [10, 100]$ m, $y_r \in [10, 100]$ m, and $z_r \in [20, 40]$ m in the simulations. For NOMA transmission, the power allocation coefficients are set as \(a_1 = 0.2\) and \(a_2 = 0.8\). In the parameter settings of the PSO algorithm, we set the swarm size to $N_p=60$ and the maximum number of iterations to $I_{\max}=100$, adopt a linearly decreasing inertia weight from $w_{\max}=0.5$ to $w_{\min}=0.4$ with $c_1=c_2=2$, limit the velocity in each dimension to $[-v_{\max},\,v_{\max}]$, and initialize all particles at $[50,50,75]^T$. To highlight the performance of FAS-RIS-NOMA networks, the FAS-RIS-OMA, TAS-RIS-NOMA and TAS-RIS-OMA are selected as the benchmarks.

\begin{figure}[h!]
    \begin{center}
        \includegraphics[width=2.9in]{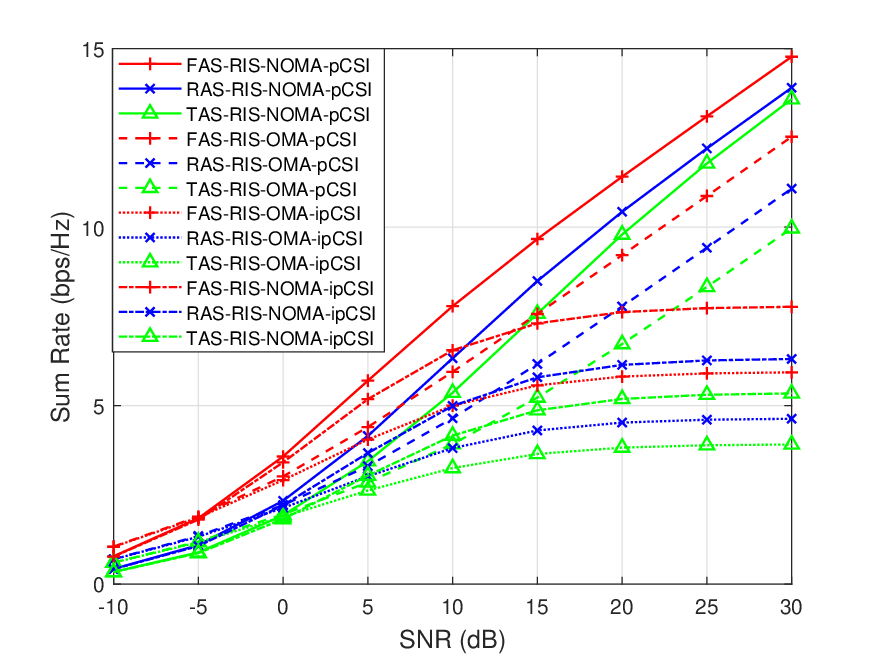}%
        \caption{Sum rate versus different values of SNR.}
        \label{NOMA-OMA}
           \vspace{-8pt}
    \end{center}
\end{figure}
Fig. 1 compares the sum rate of different antenna configurations under the NOMA and OMA schemes, where RAS denotes random port FAS. We can observe that FAS-RIS-NOMA consistently achieves the highest sum rate over the entire SNR range, mainly due to the high beamforming flexibility of FAS, which enables a more effective exploitation of spatial diversity. Benefiting from intrinsic resource multiplexing, NOMA always outperforms OMA. Additionally, we further consider the ipCSI case. Due to channel estimation errors, the sum rate growth becomes limited in the high SNR regime, resulting in a more pronounced saturation behavior. Although RAS can't perform optimal port selection, its randomly accessed port can still exploit the spatial sampling freedom within the FAS aperture, and therefore outperform TAS on average.

 \begin{figure}[!h]
    \begin{center}
        \includegraphics[width=2.9in]{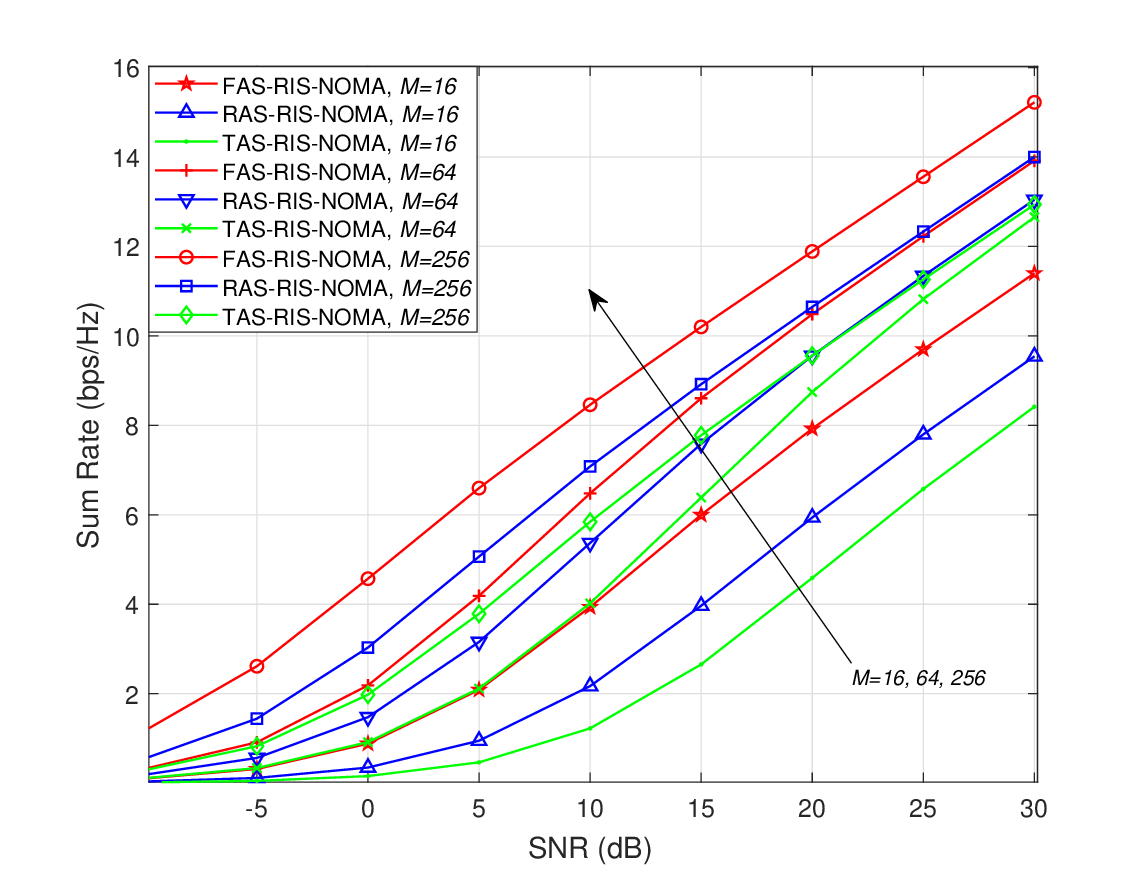}%
            \caption{Sum rate versus different values of $M$.}
        \label{M-16-64-256}
        \vspace{-8pt}
    \end{center}
\end{figure} 

Fig. 2 illustrates the impact of the number of reflecting elements on the sum rate for FAS-RIS-NOMA networks. One can seen from this figure that as the number of reflecting elements increases, the sum rate of FAS-RIS-NOMA networks is improved effectively. We can further observe that the increased number of reflectors grants the RIS greater spatial freedom, which enables more precise alignment of user channel phases while effectively compensating for path loss, ultimately leading to improved overall channel conditions for users. Additionally, increasing the number of RIS elements enhances multipath effects and further highlights the spatial diversity advantages of FAS-RIS-NOMA networks.  

 \begin{figure}[h!]
    \begin{center}
        \includegraphics[width=2.9in]{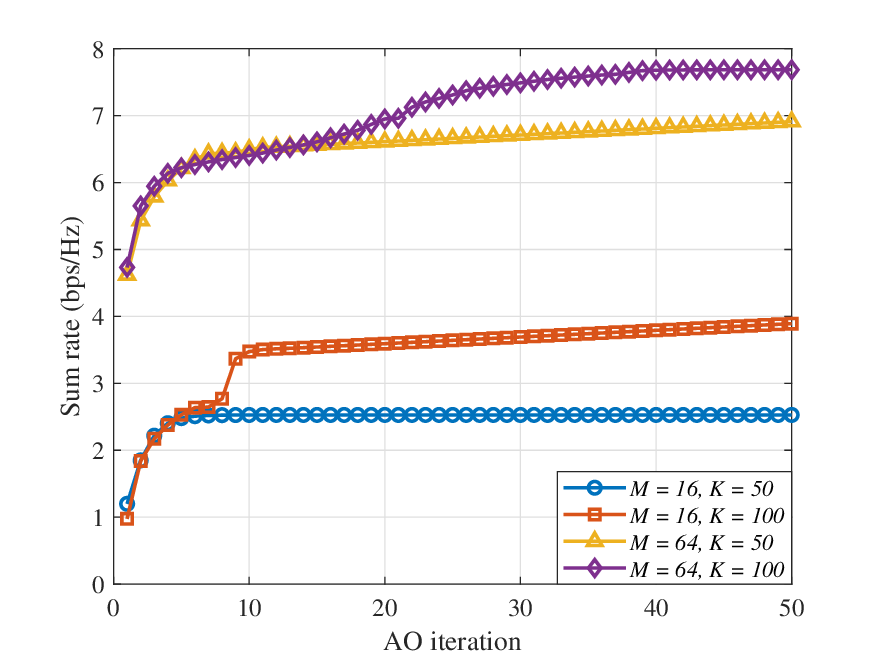}%
        \caption{Sum rate versus iteration for different $M$ and $K$.}
        \label{M-16-64-256}
         \vspace{-15pt}
    \end{center}
\end{figure} 

Fig. 3 illustrates the convergence behavior of the proposed AO algorithm under different values of $M$ and $K$. It can be observed that all curves converge rapidly within a small number of iterations, which demonstrates the good convergence and stability of the proposed algorithm. Moreover, a larger $M$ leads to a higher sum rate, since more RIS reflecting elements can provide stronger beamforming gain and improve the cascaded channel quality, thereby enhancing the sum rate. For different values of $K$, when the curves overlap, it indicates that the algorithm converges to the same optimal port selection, and thus additional candidate ports do not bring further performance gain. In contrast, when the curves are separated, a larger port search space enables the algorithm to identify a better port selection, thereby achieving a higher sum rate of FAS-RIS-NOMA networks.

\begin{figure}[h!]
    \begin{center}
        \includegraphics[width=2.9in]{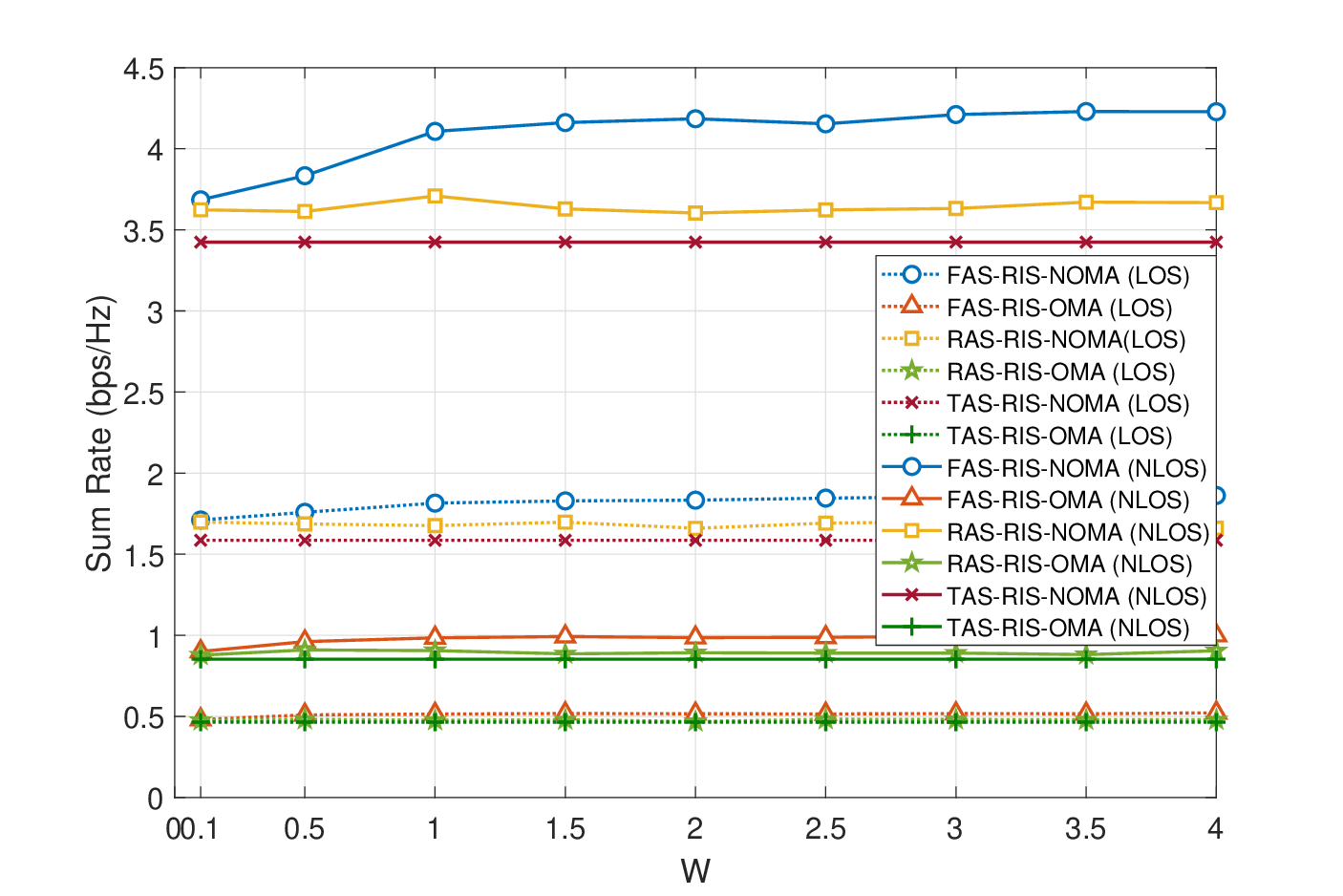}%
        \caption{Sum rate versus the different sizes of FAS.}
                \vspace{-8pt}
        \label{diff_W} 
    \end{center}
\end{figure}

Fig. 4 demonstrates the impact of FAS size on system performance under LoS and NLoS conditions. As can be observed from the simulation results that with increasing the size of FAS, the sum rate of FAS-RIS-NOMA networks improves, but the resulting stronger correlation between antenna ports limits further gains. As a result of spatial diversity from multipath propagation, the NLoS scenario outperforms the LoS case, resulting in improved performance and enhanced channel adaptation for the FAS-RIS-NOMA networks.

\section{Conclusion}\label{Conclusion1}
In this paper, we evaluated the sum rate performance of FAS-RIS-NOMA networks, in which the highly coupled non-convex optimization problem was formulated. An exhaustive search was employed for optimal port selection, the PSO algorithm was adopted to solve the RIS deployment problem, while the SCA and SDR methods were utilized to optimize the RIS phase shift matrix. Simulation results demonstrated that, benefiting from the spatial diversity provided by FAS port selection and the coherent beamforming enabled by the RIS, the FAS-RIS-NOMA network effectively enhanced the sum rate while significantly outperforming the TAS-RIS-OMA, TAS-RIS-NOMA, and FAS-RIS-OMA schemes. As the number of fluid ports increases, the port-search complexity rises significantly. Future work will consider low-complexity alternatives, such as greedy or machine learning methods, to reduce the computational overhead, and will also extend the proposed framework to more general multi-user scenarios.
\appendices
\bibliographystyle{IEEEtran}
\bibliography{reference}
\end{document}